# Assessing the Bias in Communication Networks Sampled from Twitter

Sandra González-Bailón[1], Ning Wang[1], Alejandro Rivero[2],
Javier Borge-Holthoefer[2] and Yamir Moreno[2]

Abstract

We collect and analyse messages exchanged in Twitter using two of the platform's publicly available APIs (the search and stream specifications). We assess the differences between the two samples, and compare the networks of communication reconstructed from them. The empirical context is given by political protests taking place in May 2012: we track online communication around these protests for the period of one month, and reconstruct the network of mentions and re-tweets according to the two samples. We find that the search API over-represents the more central users and does not offer an accurate picture of peripheral activity; we also find that the bias is greater for the network of mentions. We discuss the implications of this bias for the study of diffusion dynamics and collective action in the digital era, and advocate the need for more uniform sampling procedures in the study of online communication.

*Affiliations*: [1] Oxford Internet Institute, University of Oxford; [2] Institute for Biocomputation and Physics of Complex Systems, University of Zaragoza

*Corresponding author*: Sandra González-Bailón, Oxford Internet Institute, University of Oxford, 1 St. Giles, Oxford, OX1 3JS, telephone: +44 (0) 1865 287 233, fax: +44 (0) 1865 287 211, e-mail: sandra.gonzalezbailon@oii.ox.ac.uk

*Acknowledgements*: we are grateful to the audience of the OII departmental seminar for comments and suggestions, especially to Greg Taylor, Grant Blank, and Ralph Schroeder; we also thank the OUP Fell Fund and Google for their funding support.

# Assessing the Bias in Communication Networks Sampled from Twitter

**Abstract**

We collect and analyse messages exchanged in Twitter using two of the platform's publicly available APIs (the search and stream specifications). We assess the differences between the two samples, and compare the networks of communication reconstructed from them. The empirical context is given by political protests taking place in May 2012: we track online communication around these protests for the period of one month, and reconstruct the network of mentions and re-tweets according to the two samples. We find that the search API over-represents the more central users and does not offer an accurate picture of peripheral activity; we also find that the bias is greater for the network of mentions. We discuss the implications of this bias for the study of diffusion dynamics and collective action in the digital era, and advocate the need for more uniform sampling procedures in the study of online communication.

## 1. Introduction

An increasing number of studies use Twitter data to investigate a wide range of phenomena, including information diffusion and credibility, user mobility patterns, spikes of collective attention, or trends in public sentiment (Bakshy et al., 2011; Bollen et al., 2011; Castillo et al., 2011; Cha et al., 2012; Dodds et al., 2011; Lehmann et al., 2012; Paltoglou and Thelwall, 2012; Quercia et al., 2012; Romero et al., 2011; Wu et al., 2011). This boost of attention to Twitter activity responds to the prominence of the platform as a means of public communication, and to its salience in policy discussions on issues like privacy regulation, freedom of speech or law enforcement. However, the rising attention that Twitter has received form researchers is also explained by the relatively easy access to the data facilitated by the platform: unlike other prominent social networking sites (like Facebook), Twitter is public by default and the messages exchanged through the network can be downloaded at scale through the application programming interface (API) that the platform makes available to developers and, by extension, researchers.

The type of access the API offers to the underlying database of Twitter activity has changed over the years, becoming increasingly more restrictive. Currently, there are two main channels to collect messages from Twitter: the search API, which can collect messages published during the previous week but applies a rate limit to the number of queries that can be run (that is, an unspecified maximum number of calls any IP address can make per hour[1]); and the stream API, which allows requests to remain open and pushes data as it becomes available but, depending on volume, still captures just a portion of all activity taking place in Twitter (about 1% of the 'firehose' access, or complete stream of all tweets, which currently requires a commercial partnership with the platform). The questions this paper considers are: How do the data collected through the two APIs compare to each other? And if the set of messages retrieved through the APIs is not random sample of all Twitter activity, what is the nature of the bias?

These questions are important because a bias will affect not only the messages returned by the API queries but also the networks of communication that can be reconstructed from those sampled messages. Twitter activity often results from direct communication between users, either in the form of mentions or replies to other users or via the rebroadcasting of messages previously sent by someone else. Messages left out from the sample will dent the reconstruction of these communication channels and, consequently, of the emerging networks. This creates a second-order bias that can influence the answers we can give to questions of increasing interest in social science research, for instance how online networks co-evolve with offline political events and behaviour, including mass mobilisations that emerge with the support of digital media (Farell, 2012). If communication networks are reconstructed on the basis of biased samples of messages – i.e. because the original pool of messages retrieved with the API is not a random collection of all activity – then their analysis might lead to misleading conclusions about how online networks create contexts for political behaviour. The aim of this paper is to provide some evidence that can help us assess whether an identifiable bias exists and, if so, determine its nature and consequences.

[1] According to the Twitter developers page "the Search Rate Limit isn't made public to discourage unnecessary search usage and abuse" (see https://dev.twitter.com/docs/rate-limiting, accessed in November 2012).

## 2. Previous Research and Sampling Strategies

Since the launch of Twitter in 2006, an increasing body of research has tried to identify the topological properties of this communication network (Huberman et al., 2009; Java et al., 2007; Kwak et al., 2010), the position and characteristics of influential users (Bakshy et al., 2011; Cha et al., 2010), the dynamics of information exchange (boyd et al., 2010; Cha et al., 2012; Gaffney, 2010; Gonçalves et al., 2011; Honey and Herring, 2009), the existence of polarisation (boyd et al., 2010; Conover et al., 2011), and how information propagates from user to user (Borge-Holthoefer et al., 2011; Jansen et al., 2009; Romero et al., 2011; Wu et al., 2011). A search to the Web of Knowledge database for articles with Twitter as main topic returns more than 800 entries, spanning research published in conference proceedings for computer science and engineering, but also in journals of communication, media, sociology and behavioural sciences. Although all these studies are concerned with how communication takes place through the online network, the diversity of sampling frames and procedures (not to mention theoretical aims) prevent a direct comparison of their findings. Table 1 summarises the characteristics of the samples used in this previous research, giving a sense of the diversity of approaches that have been employed in the past.

-- Table 1 about here –

The references in Table 1 are not an exhaustive list of all research done with Twitter data, but they are representative of the different frames that have been applied so far to analyse Twitter communication. There are two main things to highlight from this table: one is the overlapping of observation windows across studies that used different data collection strategies; this results in redundancies in the acquisition and management of data resources, and limits the comparability of findings: although some studies have the same observation window, they do not necessarily analyse the same data. The second highlight is the different manipulations to which those samples were submitted: in some cases, the papers focus on the properties of the underlying following-follower structure, measured as a global network (Kwak et al., 2010) or at the level of dyads (Takhteyev et al., 2012); in other cases, they focus on the more direct channels of communication created by @mentions and RTs, which are employed to reconstruct ties between users (Conover et al., 2011; Grabowicz et al., 2012) or cascades of information diffusion (Bakshy et al., 2011), but also to illuminate communicative practices from a more qualitative perspective (boyd et al., 2010; Honey and Herring, 2009). In addition, these studies focus on different layers of their data depending on the information domain that is relevant to their research question; for instance, some focus on the subset of messages about trending topics (Cha et al., 2010), others analyse content related to commercial products (Jansen et al., 2009), and others analyse messages labelled with specific hashtags (Borge-Holthoefer et al., 2011; Romero et al., 2011). These studies reveal that the dynamics of communication change across different information streams; however, it is difficult to determine how many of those differences result from actual use patterns as opposed to research design choices, especially those related to data acquisition. Crucially, some of these studies cannot be replicated because they had a level of data access that is not available to most researchers.

This absence of a unified approach to sampling and data management has parallels in other online means of communication, like blogs, for which the effects of applying different sampling procedures have already been discussed (Butts and Cross, 2009). With blog data, the resulting networks also change significantly depending on the sample strategy followed and, in particular, on the selection of seed blogs. The difference with Twitter is that in this platform the researcher does not have full control on how communication is sampled due to the added constraints imposed by query limits. The API policies have become more restrictive as the number of users joining the platform was increasing exponentially; as a result, the volume of activity that can be downloaded for analysis is an increasingly smaller percentage of the full stream of activity. In the analysis of blogs, the main sampling choice is the original set of seeds from where crawling starts, and all channels for information exchange are captured as long as the crawl is set up to identify and follow all links. In Twitter, by contrast, sampling usually starts with messages, which are the main entry point to the identification of users and their exchange with other users. Messages might be filtered according to some substantive criterion related to their content (i.e. politics vs entertainment), but what the researcher does not choose or control is the filter imposed by the API when returning outputs to queries.

Most researchers acknowledge that the APIs return a semi-random sample of messages, but there is no systematic account of the bias, or how it affects inferences about the communication networks they create. Access to the full stream of activity (or firehose) would provide the natural benchmark to assess the bias that derives from using the more restrictive APIs; it would allow, for instance, testing whether the effects of the bias are greater in some information domains, and whether this is related to the volume of messages or to the diversity of users contributing those messages. The firehose access, however, is out of reach for most research organisations, which makes the assessment of the bias even more relevant: it is likely to affect most researchers. The alternative to comparing samples to the full stream of information is to compare the two available API specifications: streaming and search. The first requires more infrastructure and management time; the second needs less resources but yields lower volumes of data (more, however, if data collection needs to go back in time). As Table 1 shows, these two sources of data have been consistently used in previous research, but they have never been directly compared.

The comparison across APIs is relevant from a methodogical point of view (it tests the validity of different data collection strategies) but also for more substantive or theoretical reasons. Recent events have triggered much discussion, within and outside academia, about how and why social media facilitate collective action and political uprisings (Andersen, 2011; Farell, 2012). This interest feeds back into the longer discussion about how online technologies are changing the logic of collective action (Bimber et al., 2005; Earl and Kimport, 2011; Lupia and Sin, 2003). Networks have traditionally been analysed as the main channels through which participants are recruited and a critical mass can be attained (Diani, 2003; Marwell and Oliver, 1993). This research suggests that the dynamics of information diffusion and political mobilisation are very sensitive to the structure of the underlying communication network. With online data, communication networks are often reconstructed from messages. In Twitter, ties between users are created after parsing the messages such that if user *i* mentions or re-tweets user *j*, an arc is created from *i* to *j*. The key issue is: if the sample of messages returned by the two APIs differ, so will the networks reconstructed from them; and this has theoretical implications because it affects the empirical tests that can be run to assess research on collective action in the digital era.

Digital media have their own peculiarities, and the meaning of connections in online networks is not necessarily equivalent to what ties measured using surveys represent (McAdam, 1986; McAdam and Paulsen, 1993). However, online activity can potentially improve measurement in two important respects: first, it captures the actual intensity of communication that is directly relevant to political protests; and second, it offers a richer picture of longitudinal dynamics. In the context of political communication (especially when it is oriented to activate and coordinate protests) a random sample of all Twitter activity would not be that useful; what is interesting is the subset of all messages related to the specific event, and how the volume of those messages – and the networks of information exchange they create – evolve over time. Hashtags help identify relevant messages: they are labels contributed by users as part of their updates that classify the topic domain of their messages. The selection of the keywords that are used in data collection creates the first sampling filter: queries run through the APIs usually request messages that contain some of those keywords (Borge-Holthoefer et al., 2011; Gaffney, 2010; Yardi and boyd, 2010); or, once a data set has been collected, keywords are used to identify relevant streams and filter down larger data sets (Cha et al., 2012; Conover et al., 2011; Romero et al., 2011). But again, what the researcher does not control is the output that those queries will return when executed through the two different APIs. The following sections assess the differences across those outputs.

## 3. Data

We sampled Twitter activity around the same political protests for the period 30 April to 30 May 2012. These protests were organised to celebrate the first anniversary of the Spanish ‘indignados’ or outraged movement, which erupted in 2011 to protest against spending cuts and the management of the economic crisis. The data were collected as a follow up of previous work analysing the emergence of the movement in May 2011 (González-Bailón et al., 2012; González-Bailón et al., 2011). We collected two independent samples of messages related to the protests: the first sample (A) was collected from the UK using the search API and a list of four hashtags: these include the top three hashtags (in frequency of use) according to the sample collected in 2011 plus a new hashtag created to identify content about the 2012 mobilisations (#12M15M ). The second sample (B) was collected from Spain using the stream API and the more extensive list of 70 hashtags used in the 2011 data collection. A complete list of all these keywords can be found in the Appendix. The search API used in sample A was easier to implement than the stream API used in sample B, which required more system resources and more programming work to maintain the long-run connection with the Twitter servers. While the search API is more suitable for precise keyword data collection, and benefits from being able to retrieve messages published up to one week in the past, it is, as explained above, also more restricted in the number of calls it can make to the servers. All else equal, samples collected using the search API will be smaller than samples collected using the stream API, which is more comprehensive if the researcher has a clear idea of the messages that are of interest, via a selection of hashtags or keywords.

Table 2 summarises the datasets that resulted from using these two alternative methods. Most of the activity captured in sample A is contained in the larger sample B, but the overlapping is not complete: 2.5% of the Tweets, 1% of the authors, and 1.3% of the hashtags that appear in the small sample do not appear in the large sample. The higher volume of messages captured in the large sample, on the other hand, accounts for many interactions that go unnoticed in the smaller sample; this begs the question of how the missed information impacts on the reconstruction of communication networks. This bias can adopt different forms: by eliminating users, the small sample might be over or underestimating the network centrality of those observed; and by eliminating messages, it might underestimate the actual bandwidth of communication between users, that is, the number of messages exchanged between any two users, which can be used as a proxy to the strength of connections or the diversity of information exchanged (Aral and Van Alstyne, 2011). Given the practice of using more than one hashtag in the same message, the list of unique hashtags contained in the samples grows considerably out of the original lists used to collect the data. The overlap is close to complete (14,137 appear in both sets of messages), but the relatively smaller subset of hashtags found in sample A suggests that there is also less diversity of topics; this matters if the content of the messages is to be used as an approximation to opinions exchanged or to the framing of the protests.

-- Table 2 about here --

On the basis of these messages, we reconstructed protest-relevant communication networks using two Twitter conventions: @mentions and RTs. The first, @mentions, target other users by explicitly mentioning their username; RTs, on the other hand, are employed to broadcast messages previously published by other users (boyd et al., 2010; Honey and Herring, 2009). The networks reconstructed using these two features offer a more accurate approximation to actual communication exchange than the underlying following/follower structure: this network offers the potential for information flow but that potential is realized differently in specific information domains, as previous research suggests (Cha et al., 2010; Grabowicz et al., 2012; Romero et al., 2011). To reconstruct the @mentions and RTs networks, we parsed the text of the messages captured in the two samples and created a directed link between users *i* and *j* every time *i* mentioned or re-tweeted *j*. The resulting networks are directed and weighted, and connections are not necessarily reciprocated. Table 3 summarises the descriptive statistics for these two networks as reconstructed from the two samples. Both the @mentions and RTs networks contain about 90% of the users in a single giant connected component; the degree distribution is (in line with most online networks) very skewed, with a tail that is especially long for indegree, but the mean path length is relatively short (and similar) for all networks; they are also dissassortative, with a tendency of central users to interact with peripheral users.

-- Table 3 about here --

The vertices in these networks are subsets of the authors captured in the original sample of messages: many users sending protest messages did not engage in direct interaction with other users (via mentions or RTs), so they are excluded from the networks; but some other users present in these networks were not in the original list of sampled messages, and were identified during the snowball that resulted from parsing the messages. As Figure 1 illustrates, the percentage of users added to the @mention network during the snowball phase is nearly double for the large sample (B). These users are less central in the reconstructed network than users captured in the initial sample of messages, which means that they are less visible because they are mentioned less often. In the RTs network, very few users were snowballed, but these are again more peripheral. Figure 1 suggests that there are differences in the way users employ the @mention and RT features in the context of political protest: while the latter seem to be restricted to the most active users (who, by sending more messages, are more likely to be captured by the initial sample), mentions are used to target users that are more peripheral in this stream of information; the fact that they are targeted nonetheless suggests that they might be visible in other domains (or more central in the underlying following-follower network), which makes them be perceived as potential diffusers of information (and hence, targeted more often). This interpretation is consistent with qualitative insights on how these conventions are generally used in Twitter (boyd et al., 2010; Honey and Herring, 2009), but on the basis of observational data it is difficult to separate the confounding motivations behind their use.

-- Figure 1 about here --

Figure 2 shows that most hashtags captured by the two samples are used just once, but there are a few outliers that appear in most messages. The two most often used hashtags, ‘12M15M’ and ‘15M’ (which refer to the dates of the demonstrations for 2012 and 2011) appear in close to three quarters of all messages captured by the two samples. Most of the other hashtags co-appeared with these two main keywords. As the right panel of the figure shows, in most cases the small sample underrepresents the prominence of hashtags; but in the lower tail of the distribution the large sample also underestimates the actual frequency counts of some of these keywords. In general, however, the top 2 hashtags would have yielded the vast majority of the messages, which builds a case for restricting queries to a smaller subset of keywords – these can be identified using, for instance, trending topics; this approach will return a similar sample of messages than using a more exhaustive list of hashtags. However, for cases where the research question involves reconstructing co-occurrence networks of hashtags as a way of identifying clusters of topics (Mathiesen et al., 2012), what Figure 2 suggests is that the two APIs would add a different type of noise to the data.

-- Figure 2 about here --

## 4. Differences in Network Structure and Change

The previous section showed that the two samples allow reconstructing networks that, on the aggregate, share structural features and contain similar information. This section is concerned with individual positions within the networks, and changes over time: Do users have comparable network positions in the two samples or do these samples give us different pictures of what happens at the local level? And does the similarity of networks collected using different APIs increase or decrease over time? In order to address longitudinal changes, we reconstructed daily networks for both @mentions and RTs, and calculated the Jaccard coefficient comparing the composition of those networks according to the two samples; that is, we calculated the ratio of overlapping users over the union of all users. Figure 3 tracks changes in these coefficients; the lower panel displays changes in network sizes for reference. The figure shows that the networks grow increasingly dissimilar as time goes by, especially after the demonstrations had taken place. The number of overlapping users is maximised around the main demonstration days (May 12 and May 15), but it is also high at the pre-demonstration stage – with the exception of May 8, which coincides with a larger difference in the size of the networks (this gap, however, responds to no obvious reason).

-- Figure 3 about here --

What these trends suggest is that the picture that the two samples give of the protesters (or, to be more specific, of users engaging in the exchange and diffusion of protest-related information) is, overall, quite different – the coefficients never go much higher than 0.5; but it also means that the resolution of this picture changes over time, becoming less accurate when there is a lower volume of information exchange, as it happened after the demonstrations. This suggests that the search API, which has higher limitations in data collection, might be a less appropriate tool for periods of low activity or information streams that are not very abundant. Interestingly, though, the amount of similarity is roughly the same for the @mentions and RTs networks, although prior to the demonstration days similarity was higher for RTs. This indicates that most differences are generated by peripheral users: as mentioned above, users that remain in the smaller RTs networks are more central in the @mentions network than those who are filtered out; it is these peripheral users who make the mention networks look more different. Put differently, the fact that RTs networks are slightly more similar means that the search API returns better data for more central users, and more clustered regions of the network.

This interpretation is reinforced by the analysis of local centrality measures, summarised in Figure 4. The scatterplots correlate different measures of centrality for the users that appear in the two samples; these measures (Bonacich, 1987; Seidman, 1983) were calculated using the networks aggregated for the full observation period. The figure shows that there is a clear association of individual centrality scores, especially for the higher ranks of the distributions: in the case of degree, these are users with roughly a hundred or more connections. The k-core shows also a monotonic association, which means that the closer a user is to the core of the network according to the large sample, the closer the same user is to the core of the network according to the small sample. However, for users that are less well connected, the small sample underestimates centrality more clearly. This is related to the observation made above that larger samples are better at getting messages that target or re-tweet to or from the peripheral users. These messages are absent from the small sample and, as a result, undervalue the centrality of many users.

Centrality in this context means that someone is mentioned more often or re-tweeted more times in the flow of protest-related communication. It is a direct measure of visibility in this information domain, but also of outreach: it helps identify who is being more active in promoting protest-related talk. The disagreement between samples on the centrality scores of many users has different implications depending on the research question at hand. If the topic of interest is who are the leaders of the mobilisation (in terms of who is seeding the network with relevant information) then both samples would give a similar picture of the users that populate the core of activity. But if the question is how that information spreads, given that most users are peripheral and that this is the subset of nodes where most disagreement takes place, conclusions should be more cautious, especially for the smaller sample. Here the notions of centre and periphery relate to the density of connections, and to the extent to which the network has a class of actors that are loosely connected to a more cohesive subgraph (Borgatti and Everett, 1999): the existence of a core/periphery structure in networks does not necessarily imply a hierarchical organisation. What the findings above suggest is that the search API is worse at tapping those peripheral, loosely connected users – who happen to be the majority in online networks. The extra information captured by the streaming API lies mostly at the fringes of the communication network.

-- Figure 4 about here --

## 5. Measuring the Bias

The analyses above suggest that the smaller sample (A) is not a random subset of the larger sample (B); this, in turn, suggests that sample B is probably not a random subset of the full stream of protest messages either, but to the extent that the full stream is not available for research, we can only speculate about the extent of the departure. Instead, we can measure the bias in the small sample by comparing the networks that result from it with the networks that would result from a random selection of nodes in the large sample. The process consists of iterating a random selection process so that in every round a subgraph is extracted from the large network by randomly selecting N vertices (and the adjacent edges), where N equals the size of the small @mention and RTs networks. This is repeated 1000 times. Network statistics are then calculated for the thousand random networks, which helps build a theoretical probability distribution with which to assess the magnitude of the observed statistics, and the extent to which they depart from what would be expected by random chance. This comparison makes sense to the extent that the overlap between samples is close to complete.

Figure 5 shows the results of this iteration process for a few network statistics. Starting with the @mention network, the boxplots indicate that the small sample significantly overestimates degree centralisation, which is close to 50% -- none of the random networks yield values as high for degree centralisation, a departure that also applies to the maximum k-shell: the observed network has a much more cohesive core than any of the random networks. As a result, the simulated networks also exhibit a significantly higher number of components and higher average path length. Differences are smaller for local density and degree correlation, although the observed network exhibits a greater tendency for core-periphery interactions than most simulated networks. These differences are smaller for the RTs network, but significant in some cases: the observed network is less centralised than the network formed by mentions, but it still exhibits extreme centralisation, compared to the random networks, and a more cohesive core than would be expected by chance. Degree correlation falls within the interquartile range, and so does clustering, which suggest that the bias has a lesser effect on these properties, at least on the aggregate level.

-- Figure 5 about here --

On the basis of these findings we can conclude two things: first, that networks formed by mentions are more biased towards central users than networks formed by RTs; and second, that this bias might still be underestimated if the large sample is also biased towards more central users compared to the full stream. Again, the implications of this bias will vary with the research question, but it is particularly relevant for the study of core-periphery dynamics and, in the context of online collective action, the emergence of a critical mass. Peripheral users do not have many connections, and they do not offer the important voices in terms of impact or reach; but their superiority lies in the numbers: most users qualify as peripheral (this is what the skewed degree distributions tell us) and they are the mass that is activated when political protests are successful. The bias in the data means that it is difficult to assess how large that periphery is and, by extension, how long it takes to activate them; it also undermines the study of diffusion dynamics because it shrinks the actual size of the population of interest.

## 6. Discussion

The rising interest in digital media and social interactions mediated by online technologies is boosting the research outputs in an emerging field that is multidisciplinary by nature: it brings together computer scientists, sociologists, physicists, and researchers from a wide range of other disciplines. The collaborative nature of this research is making the number of studies increase fast, often to catch up with the rise of new technologies; but the speed at which the field moves means that standards often do not have time to mature and consolidate. Procedures for data collection and sampling offer a particularly good example of this lack of consolidation: they vary widely across disciplines, and often adapt differently to the peculiarities of the platforms from where the data are gathered. As a result, it is difficult to integrate research outputs, especially when proprietary data are involved in the analyses – this hampers the ability to replicate findings and engage in cumulative research and theory building.

This paper has tackled this lack of standards by focusing on one particular platform, Twitter, currently one of the most popular SNSs (at least in Western societies), and at the centre of much public and media attention for its alleged role in protests and mass mobilisations. In particular, the paper has compared the networks of communication that can be reconstructed when two different sampling strategies are applied to the same underlying population of messages. What the comparison reveals is that there is a bias in the reconstructed networks that goes unnoticed in most research but that might have important theoretical implications for some of the questions that have been posed to the data in the past.

The virtues of online data are so great (in terms of their scale and resolution) that the problems and deficiencies are often dismissed without a systematic account of how they might interfere with the analyses. The conclusions of this paper are limited to the particular information context analysed (i.e. politically relevant communication in the context of mass mobilisations), and are subject to the limitations of not being able to access unfiltered data (at the moment of writing, this requires a commercial agreement with Twitter). But they provide enough evidence to defend the claim that a more careful account of data quality and bias, and the creation of standards that can facilitate the comparability of findings, would benefit an emerging area of research, especially if it is to yield insights that can survive specific technologies and short-sighted research aims. The Library of Congress announced in 2010 its plans to archive every tweet since 2006; however, two years later these plans have yet to fructify. In the meantime, researchers interested in reconstructing the trails of online communication will have to factor in their analysis the bias inherent to their data.

## References

Andersen, K., 2011. The Protester, Time.

Aral, S., Van Alstyne, M., 2011. The Diversity-Bandwith Trade-off. American Journal of Sociology 117, 90-171.

Bakshy, E., Hofman, J.M., Mason, W.A., Watts, D.J., 2011. Everyone's an Influencer: Quantifying Influence on Twitter, Proceeding of the Fourth International Conference on Web Search and Data Mining (WSDM 2011).

Bimber, B., Flanagin, A., Sthol, C., 2005. Reconceptualizing Collective Action in the Contemporary Media Environment. Communication Theory 15, 365-388.

Bollen, J., Mao, H., Zeng, X.-J., 2011. Twitter Mood Predicts the Stock Market. Journal of Computational Science 2, 1-8.

Bonacich, P., 1987. Power and Centrality: A family of Measures. American Journal of Sociology 92, 1170-1182.

Borgatti, S.P., Everett, M.G., 1999. Models of core/periphery structures. Social Networks 21, 375-395.

Borge-Holthoefer, J., Rivero, A., García, I., Cauhé, E., Ferrer, A., Ferrer, D., Francos, D., Iñiguez, D., Pérez, M.P., Ruiz, G., Sanz, F., Serrano, F., Viñas, C., Tarancón, A., Moreno, Y., 2011. Structural and Dynamical Patterns on Online Social Networks: the Spanish May 15th Movement as a Case Study doi:10.1371/journal.pone.0023883. PloS ONE 6, 6e23883.

boyd, d., Golder, S.A., Lotan, G., 2010. Tweet, Tweet, Retweet: Conversational Aspects of Retweeting on Twitter, HICSS-43 IEEE, Kauai, HI.

Butts, C.T., Cross, R.B., 2009. Change and External Events in Computer-Mediated Citation Networks: English Language Weblogs and the 2004 U.S. Electoral Cycle. Journal of Social Structure 10.

Castillo, C., Mendoza, M., Poblete, B., 2011. Information Credibility on Twitter, Proceedings of the 20th International World Wide Web Conference (WWW 2011).

Cha, M., Benevenuto, F., Haddadi, H., Gummadi, K.P., 2012. The World of Connections and Information Flow in Twitter. IEEE Transactions on Systems, Man and Cybernetics 42, 991-998.

Cha, M., Haddadi, H., Benevenuto, F., Gummadi, K.P., 2010. Measuring User Influence in Twitter: The Million Follower Fallacy, Proceedings of the International AAAI Conference on Weblogs and Social Media (ICWSM 2010).

Conover, M.D., Ratkiewicz, J., Francisco, M., Goncalves, B., Flammini, A., Menczer, F., 2011. Political Polarization on Twitter, International Conference on Weblogs and Social Media (ICWSM'11).

Diani, M., 2003. Networks and Social Movements: a Research Programme, in: Diani, M., McAdam, D. (Eds.), Social Movements and Networks. Relational Approaches to Collective Action. Oxford University Press, New York.

Dodds, P.S., Harris, K.D., Kloumann, I.M., Bliss, C.A., Danforth, C.M., 2011. Temporal Patterns of Happiness and Information in a Global Social Network: Hedonometrics and Twitter. PloS ONE 6, e26752.

Earl, J., Kimport, K., 2011. Digitally Enabled Social Change: Activism in the Internet Age. MIT, Cambridge, MA.

Farell, H., 2012. The Consequences of the Internet for Politics. Annual Review of Political Science 15, 35-52.

Gaffney, D., 2010. #iranElection: Quantifying Online Activism. Proceedings of the WebSci10: Extending the Frontiers of Society On-Line.

Gonçalves, B., Perra, N., Vespignani, A., 2011. Modeling Users' Activity on Twitter Networks: Validation of Dunbar's Number. PloS ONE 6, e22656.

González-Bailón, S., Borge-Holthoefer, J., Moreno, Y., 2012. Broadcasters and Hidden Influentials in Online Protest Diffusion. American Behavioral Scientist forthcoming.

González-Bailón, S., Borge-Holthoefer, J., Rivero, A., Moreno, Y., 2011. The Dynamics of Protest Recruitment through an Online Network. Scientific Reports 1.

Grabowicz, P.A., Ramasco, J.J., Moro, E., Pujol, J.M., Eguiluz, V.M., 2012. Social Features of Online Networks: The Strength of Intermediary Ties in Online Social Media. PloS ONE 7, e29358.

Honey, C., Herring, S.C., 2009. Beyond Microblogging: Conversation and Collaboration via Twitter, System Sciences, 2009. HICSS '09. 42nd Hawaii International Conference on, pp. 1-10.

Huberman, B.A., Romero, D.M., Wu, F., 2009. Social networks that matter: Twitter under the microscope. First Monday 14, http://firstmonday.org/htbin/cgiwrap/bin/ojs/index.php/fm/article/viewArticle/2317/2063.

Jansen, B.J., Zhang, M., Sobel, K., Chowdury, A., 2009. Twitter power: Tweets as electronic word of mouth. J. Am. Soc. Inf. Sci. Technol. 60, 2169-2188.

Java, A., Song, X., Finin, T., Tseng, B., 2007. Why we twitter: understanding microblogging usage and communities, Proceedings of the 9th WebKDD and 1st SNA-KDD 2007 workshop on Web mining and social network analysis. ACM, San Jose, California, pp. 56-65.

Kwak, H., Lee, C., Park, H., Moon, S., 2010. What is Twitter, a Social Network or a News Media?, Proceedings of the 19th International World Wide Web Conference (WWW 2010).

Lehmann, J., Goncalves, B., Ramasco, J.J., Catuto, C., 2012. Dynamical Classes of Collective Attention in Twitter, World Wide Web, Lyon, France.

Lupia, A., Sin, G., 2003. Which public goods are endangered? How evolving communication technologies affect *The logic of collective action*. Public Choice 117, 315-331.

Marwell, G., Oliver, P., 1993. The Critical Mass in Collective Action. Cambridge University Press, Cambridge.

Mathiesen, J., Yde, P., Jensen, M.H., 2012. Modular networks of word correlations on Twitter. Scientific Reports 2.

McAdam, D., 1986. Recruitment to High-Risk Activism: The Case of Freedom Summer. American Journal of Sociology 92, 64-90.

McAdam, D., Paulsen, R., 1993. Specifying the Relationship between Social Ties and Activism. American Journal of Sociology 99, 640-667.

Paltoglou, G., Thelwall, M., 2012. Twitter, MySpace, Digg: Unsupervised sentiment analysis in social media. ACM Transactions on Intelligent Systems and Technology (TIST) 3, 66:61-19.

Quercia, D., Capraz, L., Crowcroft, J., 2012. The SocialWorld of Twitter: Topics, Geography, and Emotions, AAAI  ICWSM, Dublin.

Romero, D.M., Meeder, B., Kleinberg, J., 2011. Differences in the Mechanics of Information Diffusion Across Topics: Idioms, Political Hashtags, and Complex Contagion on Twitter, International World Wide Web Conference, Hyderabad, India.

Seidman, S.B., 1983. Network structure and minimum degree. Social Networks 5, 269-287.

Takhteyev, Y., Gruzd, A., Wellman, B., 2012. Geography of Twitter Networks. Social Networks 34, 73-81.

Wu, S., Hofman, J.M., Mason, W.A., Watts, D.J., 2011. Who Says What to Whom on Twitter, Proceedings of the 20th International World Wide Web Conference (WWW 2011).

Yardi, S., boyd, d., 2010. Dynamic Debates: An Analysis of Group Polarization over Time on Twitter. Bulletin of Science, Technology & Society 30, 316-327.

Appendix

Table A1. List of 70 hashtags used in API queries (ranked by frequency of use)

| Rank | Hashtag |
|---|---|
| 1 | #acampadasol |
| 2 | #spanishrevolution |
| 3 | #nolesvotes |
| 4 | #15m |
| 5 | #nonosvamos |
| 6 | #democraciarealya |
| 7 | #notenemosmiedo |
| 8 | #yeswecamp |
| 9 | #15mani |
| 10 | #acampadasevilla |
| 11 | #globalcamp |
| 12 | #acampadavalencia |
| 13 | #acampadagranada |
| 14 | #acampadamalaga |
| 15 | #acampadazgz |
| 16 | #consensodeminimos |
| 17 | #italianrevolution |
| 18 | #estonosepara |
| 19 | #acampadaalicante |
| 20 | #tomalacalle |
| 21 | #europeanrevolution |
| 22 | #acampadapamplona |
| 23 | #worldrevolution |
| 24 | #acampadapalma |
| 25 | #tomalaplaza |
| 26 | #acampadas |
| 27 | #15mpasalo |
| 28 | #cabemostodas |
| 29 | #nonosmovemos |
| 30 | #3puntosbasicos |
| 31 | #frenchrevolution |
| 32 | #estonoseacaba |
| 33 | #acampadatoledo |
| 34 | #nonosrepresentan |
| 35 | #acampadalondres |
| 36 | #globalrevolution |
| 37 | #acampadazaragoza |
| 38 | #acampadaparis |
| 39 | #takethesquare |
| 40 | #periodismoeticoya |
| 41 | #hastalasgenerales |
| 42 | #irishrevolution |
| 43 | #democraziarealeora |
| 44 | #democraciaparticipativa |

| Rank | Hashtag |
|---|---|
| 45 | #15mpamplona |
| 46 | #barcelonarealya |
| 47 | #dry_jaen |
| 48 | #usarevolution |
| 49 | #dry_caceres |
| 50 | #dryasturies |
| 51 | #democraziareale |
| 52 | #democratiereelle |
| 53 | #dry_cadiz |
| 54 | #dry_toledo |
| 55 | #acampadasvlla |
| 56 | #drybizkaia |
| 57 | #dry_santander |
| 58 | #15mayovalencia |
| 59 | #dry_pisa |
| 60 | #dryginebra |
| 61 | #DRY_Algeciras |
| 62 | #demorealyaib |
| 63 | #DRYGipuzkoa |
| 64 | #DryValladolid |
| 65 | #ItalRevolution |
| 66 | #BolognaDRY |
| 67 | #DRY_Pavia |
| 68 | #DRY_Almeria |
| 69 | #15mayoCordoba |
| 70 | #ciudades-dry |

Table 1. Sample Characteristics of Previous Research on Twitter

| Reference | Number of messages | Observation window (DD/MM/YY) | Source* | Networks? | | |
|---|---|---|---|---|---|---|
| | | | | following | @mentions | RTs |
| 2007 Java et al. | 1.5 million | 01/04/07 to 30/05/07 | public timeline | Y | N | N |
| 2008 Huberman et al. | 79 million | not specified | search API | Y | Y | N |
| 2009 Jansen et al. | 150 thousand | 04/04/08 to 03/07/08 | search API | N | N | N |
| 2010 boyd et al. | 720 thousand<br>203 thousand | 26/01/09 to 13/06/09<br>20/04/09 to 13/06/09 | public timeline<br>search API | N<br>N | N<br>N | Y<br>Y |
| 2010 Cha et al. | 1.8 billion | 01/05/2006 to 30/08/2009** | search API (white-listed) | Y | Y | Y |
| 2010 Gaffney | 770 thousand | 16/06/2009 to 23/10/2009 | search API (white-listed) | N | N | Y |
| 2010 Kwak et al. | 106 million | 06/06/2009 to 31/06/2009 | search API (white-listed) | Y | N | Y |
| 2010 Yardi et al. | 30 thousand | 31/05/2009 to 01/06/2009 | search API (white-listed) | N | N | N |
| 2011 Bakshy et al. | 1.03 billion | 13/09/2009 to 15/11/2009 | streaming API (firehose) | Y | N | Y |
| 2011 Borge-Holthoefer et al. | 189 thousand | 25/04/2011 to 26/05/2011 | streaming API | N | Y | N |
| 2011 Conover et al. | 250 thousand | 14/09/2010 to 01/11/2010 | streaming API (gardenhose) | N | Y | Y |
| 2011 Gonçalves et al. | 382 million | 01/05/2006 to 30/05/09** | streaming API (firehose) | Y | Y | N |
| 2011 Romero et al. | 3 billion | 01/08/2009 to 01/01/2010 | search API | N | Y | N |

Table 1. Sample Characteristics of Previous Research on Twitter (continued)

| Reference | Number of messages | Observation window (DD/MM/YY) | Source* | Networks? | | |
|---|---|---|---|---|---|---|
| | | | | following | @mentions | RTs |
| 2011 Wu et al. | 5 billion | 28/07/2009 to 08/03/2010 | streaming API (firehose) | Y | N | Y |
| 2012 Takhteyev et al. | 481 thousand | 01/08/2009 to 07/08/2009** | public timeline | Y | N | N |
| 2012 Grabowicz et al. | 12 million | 20/11/2008 to 11/12/2008 | search API (white-listed) | Y | Y | Y |

Notes: * 'firehose' gives access to the full stream of messages (it currently requires commercial agreement); the free access stream API (a.k.a. 'spritz') returns about 1% of the full firehose; 'gardenhose' access, now discontinued, returned about 10% of the stream. White-listed accounts could run up to 20k queries per hour using the search API (white-listing is no longer provided). **Actual dates of data collection not specified (only months or years).

Table 2. Size of the Datasets collected using Search (sample A) and Stream (sample B) APIs

| | Sample A | Sample B | Overlap |
|---|---|---|---|
| Number of Tweets | 272,944 | 1,026,291 | 266,123 |
| Number of unique authors | 71,285 | 217,284 | 70,605 |
| Number of unique hashtags | 14,324 | 46,546 | 14,137 |

Table 3. Descriptive Statistics for the @mention and RTs Networks

| | Small Sample (A) | | Large Sample (B) | |
|---|---|---|---|---|
| | @mentions | RTs | @mentions | RTs |
| *N* (number of vertices) | 68,196 | 56,196 | 222,017 | 163,873 |
| *M* (number of edges) | 200,039 | 121,492 | 706,176 | 450,857 |
| $<k>$ (mean degree) | 6 | 4 | 6 | 6 |
| max($k_{in}$) (maximum indegree) | 5,204 | 3,930 | 13,420 | 11,460 |
| max($k_{out}$) (maximum outdegree) | 619 | 259 | 2,817 | 450 |
| *l* (mean path length) | 5.18 | 5.17 | 5.32 | 5.53 |
| *C* (clustering) | 0.014 | 0.008 | 0.012 | 0.009 |
| *r* (degree correlation coefficient) | -0.098 | -0.097 | -0.095 | -0.089 |
| # strong components | 60,857 | 52,816 | 198,223 | 151,566 |
| *N* giant component | 62,403 (92%) | 51,037 (91%) | 198,587 (90%) | 149,070 (91%) |
| *N* $2^{nd}$ component | 2 | 2 | 5 | 2 |

Figure 1. Users Aggregated to the Networks at Different Stages of Data Collection

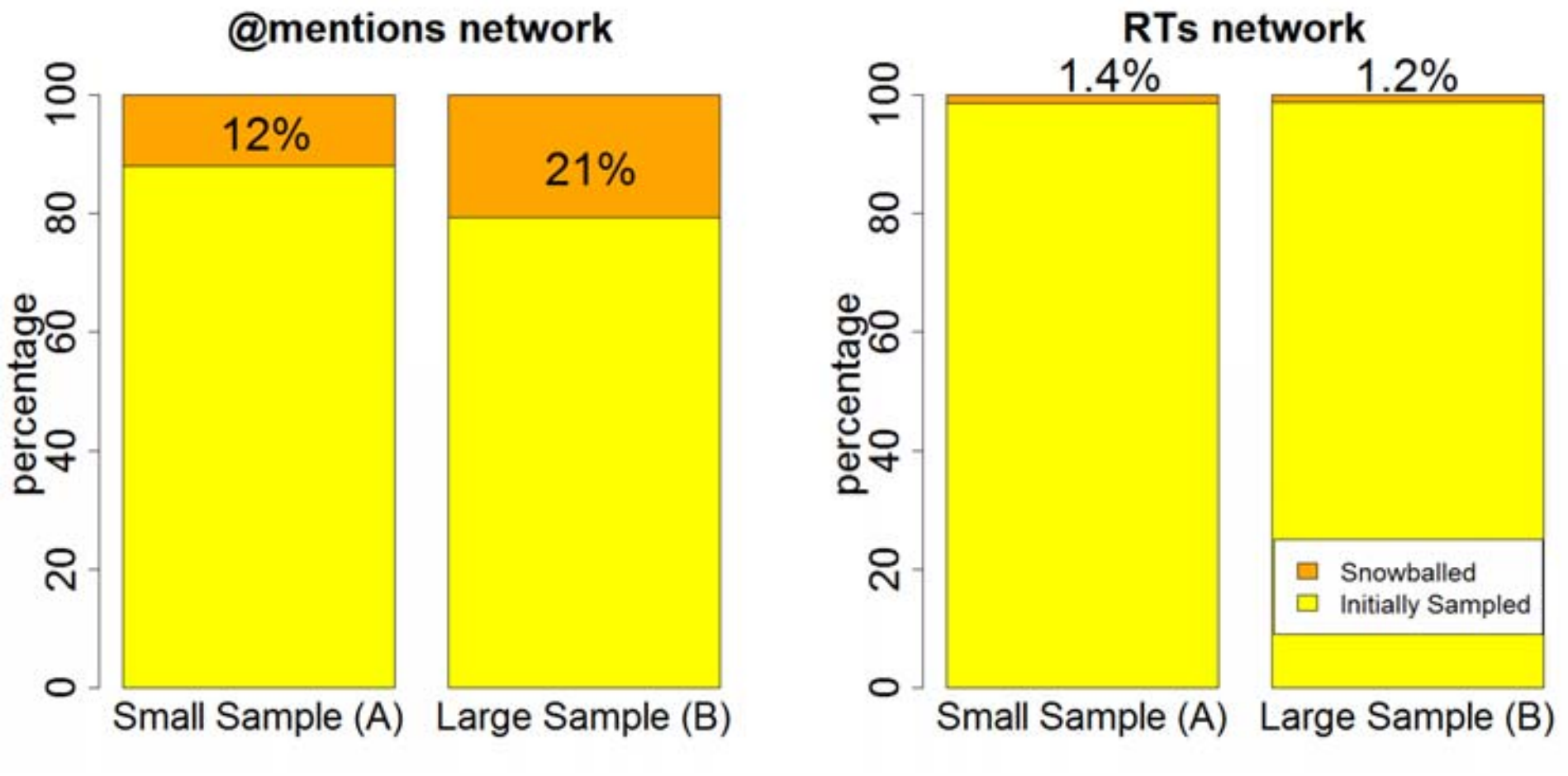

Figure 2. Frequency of Use of Hashtags

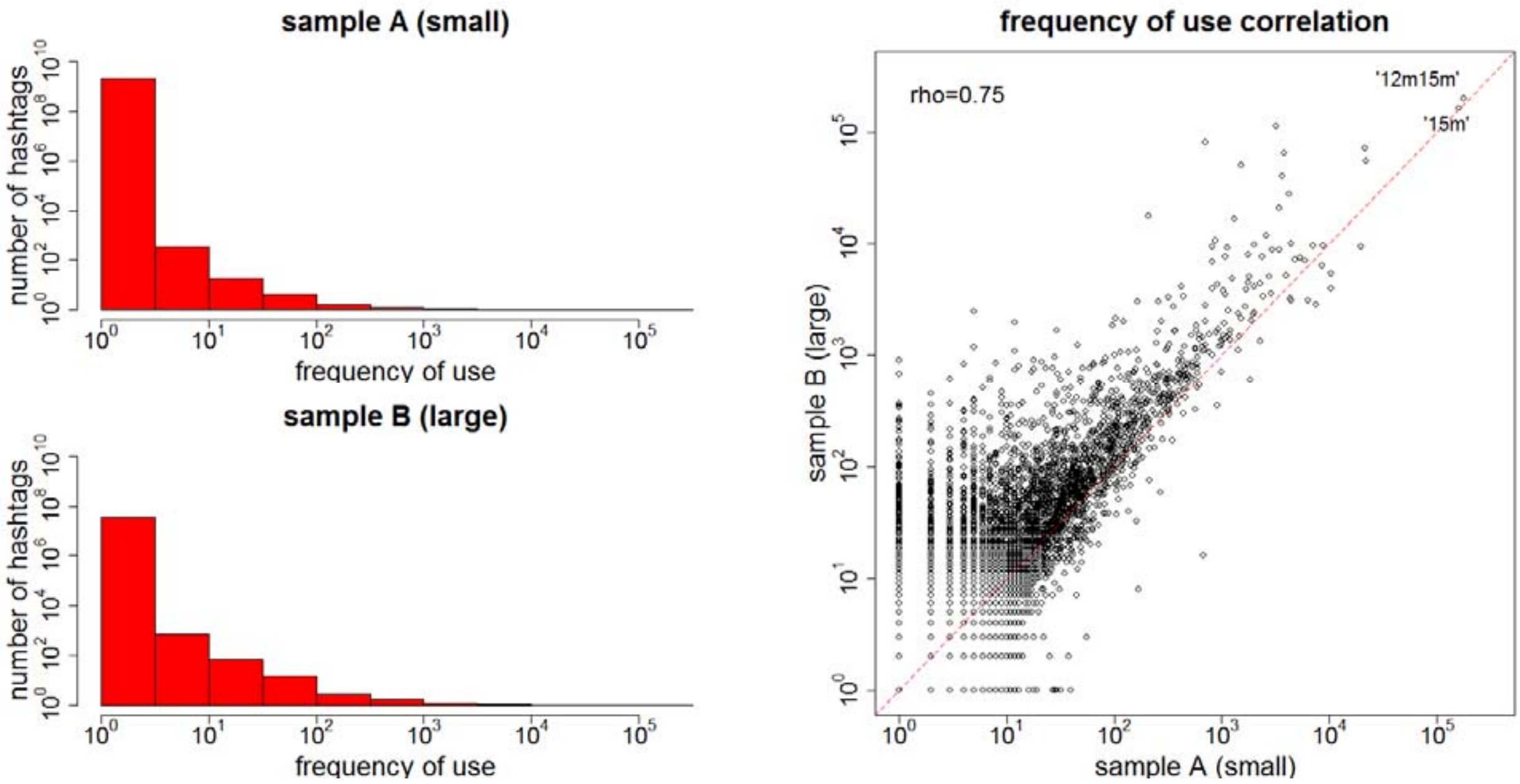

Figure 3. Similarity of Networks Reconstructed using the Two Samples of Messages

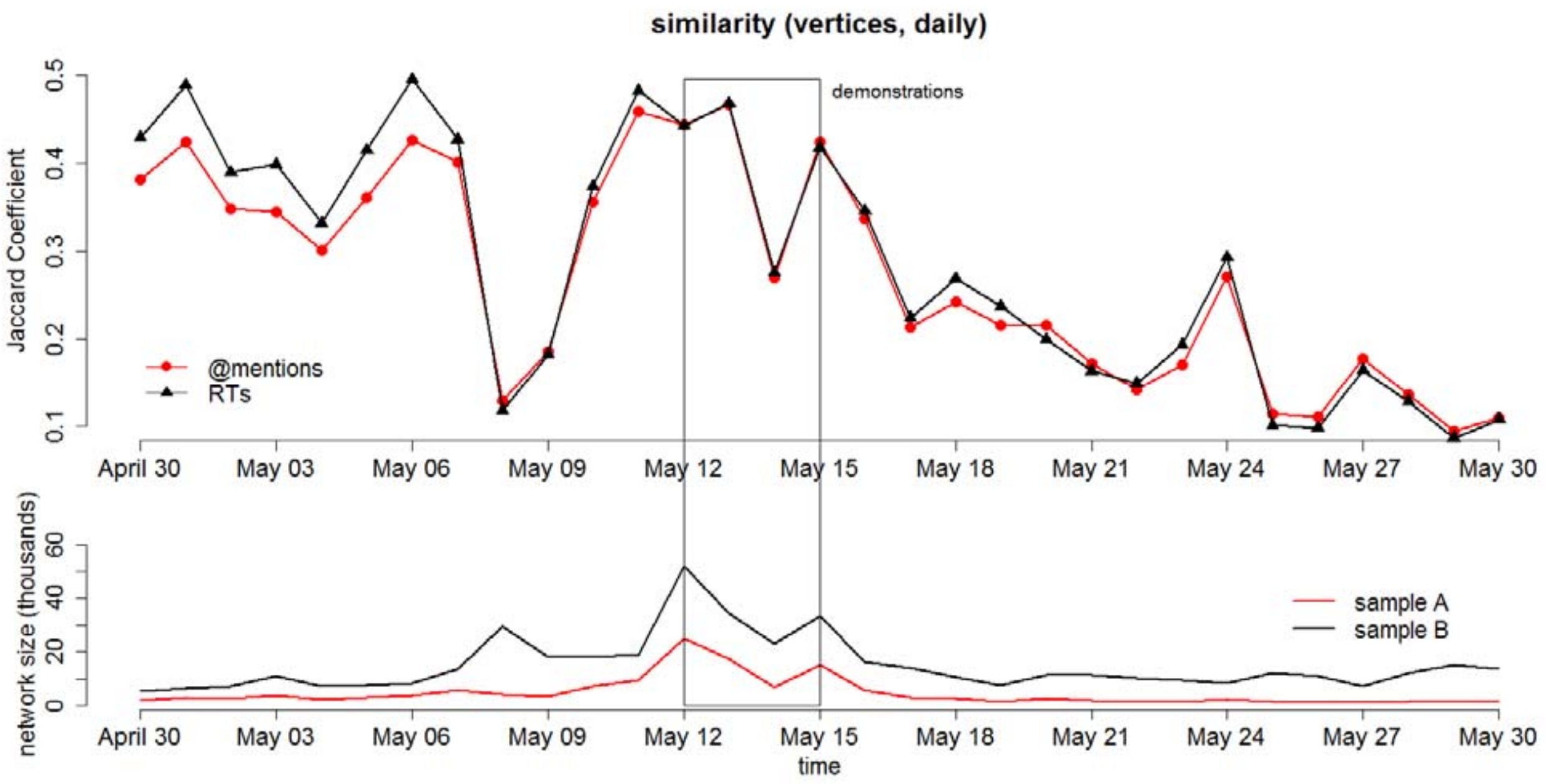

Figure 4. Correlation of Centrality Measures in both Samples

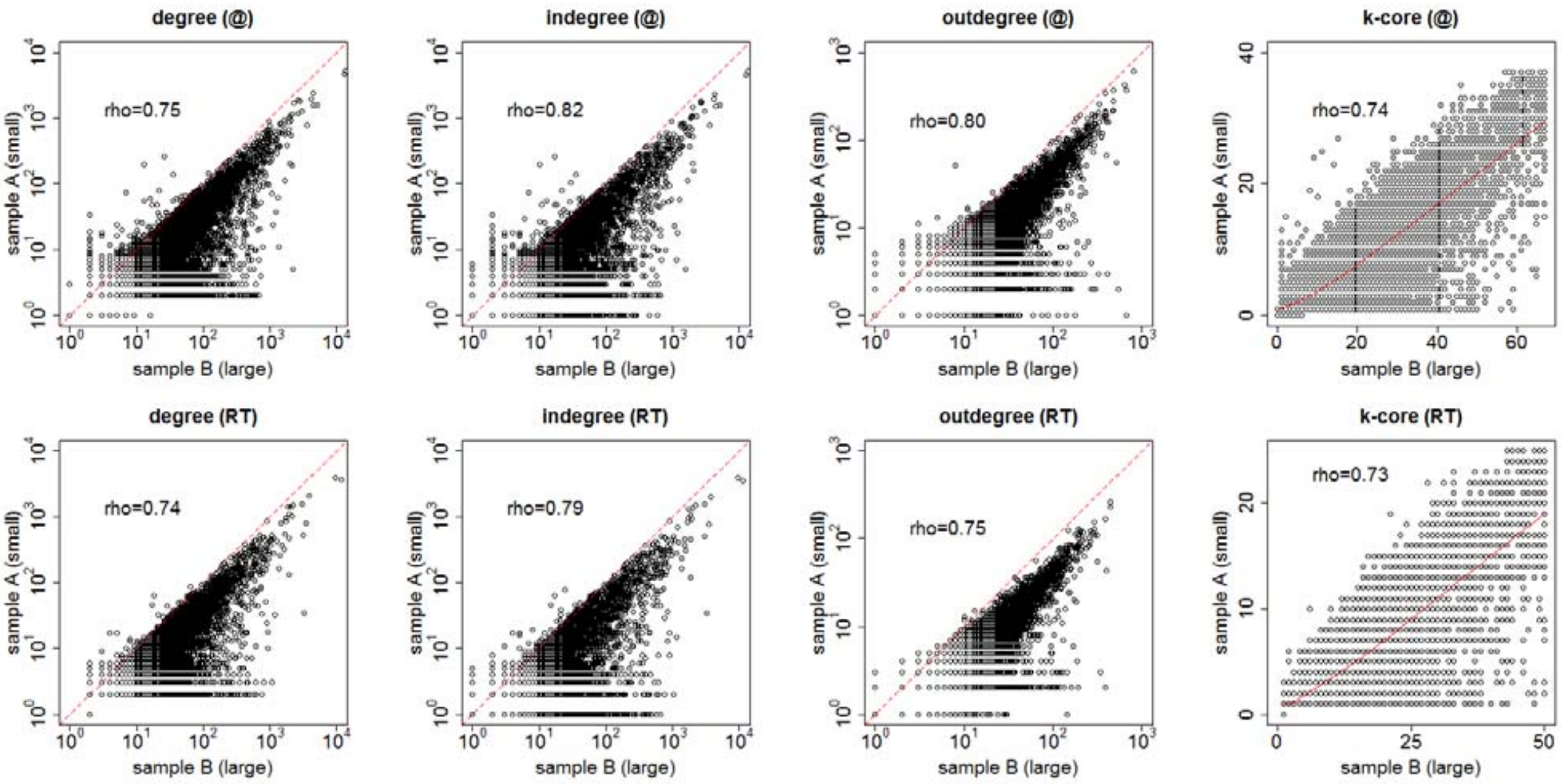

Figure 5. Observed Statistics for Small Networks (red dots) compared to Random Draws from the Large Sample (1000 subgraphs)

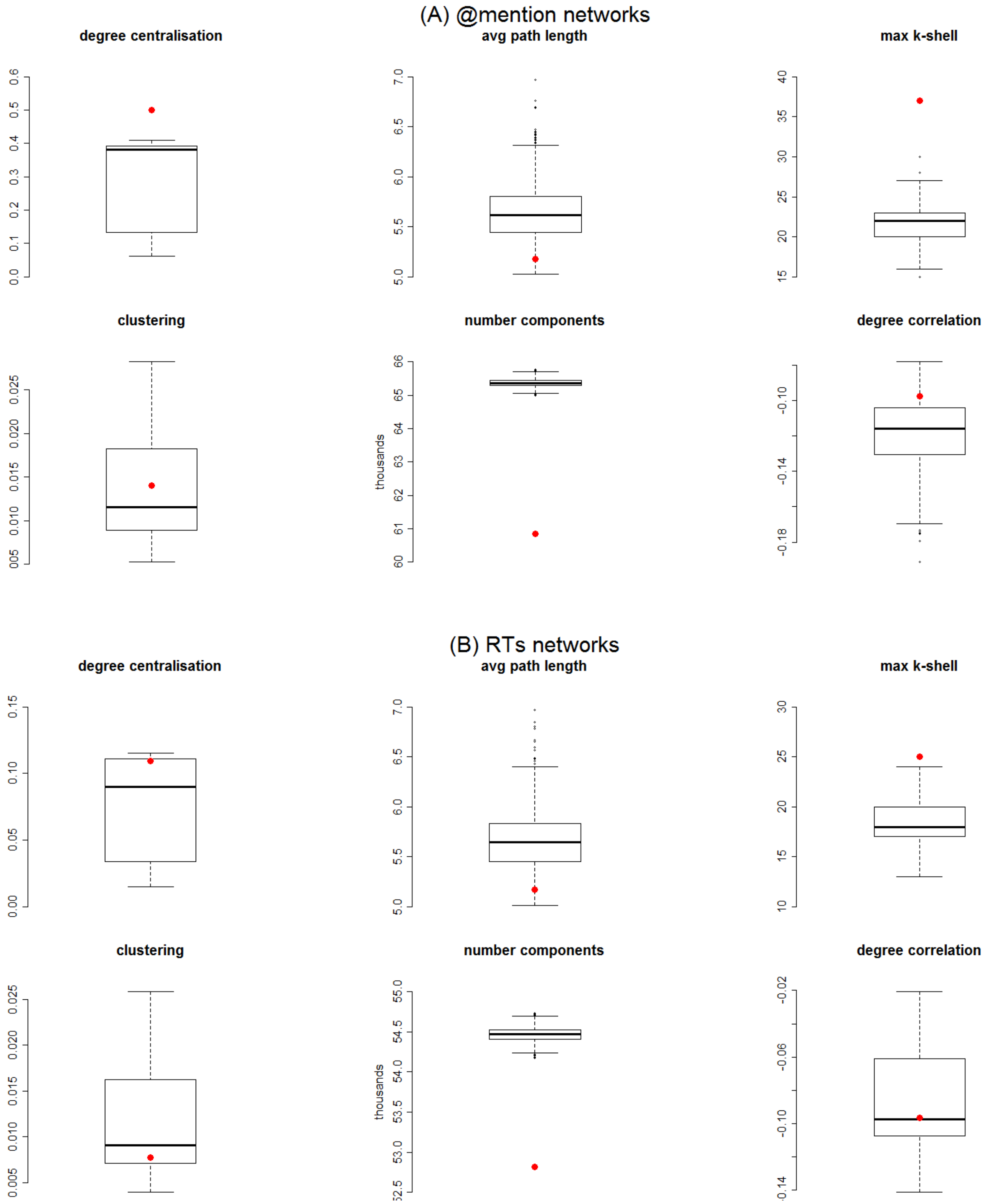